\documentclass[fleqn,10pt]{wlscirep}
\usepackage[utf8]{inputenc}
\usepackage[T1]{fontenc}
\usepackage{amsmath}
\usepackage{graphicx}
\usepackage{booktabs}

\begin{document}


\title{First-principles Dzyaloshinskii-Moriya interaction in a non-collinear framework }

\author[1,2,3,*]{R. Cardias}
\author[2]{A. Szilva}
\author[4]{M. M. Bezerra-Neto}
\author[5]{M. S. Ribeiro}
\author[2]{A. Bergman}
\author[2]{Y. O. Kvashnin}
\author[2]{J. Fransson}
\author[1]{A. B. Klautau}
\author[2,6]{O. Eriksson}
\author[2]{L. Nordstr\"om}

\affil[1]{Faculdade de F\'\i sica, Universidade
Federal do Par\'a, Bel\'em, PA, Brazil}
\affil[2]{Department of Physics and Astronomy, Uppsala University, 75120 Box 516 Sweden}
\affil[3]{SPEC, CEA, CNRS, Université Paris-Saclay, CEA Saclay, 91191 Gif-Sur-Yvette, France}
\affil[4]{Instituto de Engenharia e Geoci\^encias, Universidade Federal do Oeste do Par\'a, Santar\'em, PA, Brazil}
\affil[5]{Instituto Federal do Par\'a, Campus Bel\'em, PA, Brazil}
\affil[6]{School of Science and Technology, \"Orebro University, SE-701 82 \"Orebro, Sweden}
\affil[*]{ramon.cardias@cea.fr}

\date{\today}

\begin{abstract}

\noindent

We have derived an expression of the Dzyaloshinskii-Moriya (DM) interaction, where all the three components of the DM vector can be calculated independently, for a general, non-collinear magnetic configuration. The formalism is implemented in a real space – linear muffin-tin orbital – atomic sphere approximation (RS-LMTO-ASA) method. We have chosen the Cr triangular trimer on Au(111) and Mn triangular trimers on Ag(111) and Au(111) surfaces as numerical examples. The results show that the DM interaction (module and direction) is drastically different for collinear and non-collinear states. Based on the relation between the spin and charge currents flowing in the system and their coupling to the non-collinear magnetic configuration of the triangular trimer, we demonstrate that the DM interaction can be significant, even in the absence of spin-orbit coupling. This is shown to emanate from the non-collinear magnetic structure, that can induce significant spin and charge currents even with spin-orbit coupling is ignored.


\end{abstract}
\maketitle
%
\section*{Introduction}

Chiral objects are unique and present intriguing properties. In the field of magnetism, topologically quantized magnetic whirl textures, called Skyrmions, have been widely explored in literature \cite{heinzeSpontaneousAtomicscaleMagnetic2011,pereiroTopologicalExcitationsKagome2014,fertMagneticSkyrmionsAdvances2017,hoshinoTheoryMagneticSkyrmion2018,zhangMagneticSkyrmionsSkyrmion2017,maccarielloElectricalDetectionSingle2018,seabergNanosecondXRayPhoton2017,matsumotoStableMagneticSkyrmion2018,jalilStabilityTopologicalCharge2016}. Due to its topological protection, the skyrmion is regarded as a novel particle with potential applicability in information technology. Chiral magnets are characterized by the presence of the Dzyaloshinskii-Moriya interaction (DMI)\cite{dzyaloshinskyThermodynamicTheoryWeak1958,udvardiFirstprinciplesRelativisticStudy2003,moriyaAnisotropicSuperexchangeInteraction1960}, which results from spin-orbit coupling (SOC) combined with absence of inversion symmetry \cite{udvardiFirstprinciplesRelativisticStudy2003}.  Both strength and direction of the DMI have been reported\cite{crepieuxDzyaloshinskyMoriyaInteractions1998,antalFirstprinciplesCalculationsSpin2008,udvardiChiralAsymmetrySpinWave2009}, where the latter is known due to symmetry reasons for the cases of bulk, surfaces and even impurities on surfaces;  but to calculate the DMI for nanostructures with low symmetry it is still a challenge. 

Several methods in the literature are used to study statical and dynamical properties of magnetic materials, such as magnetic ground state, phase transition, critical temperature, magnetic dynamics, etc. In order to achieve that, spin-Hamiltonians can be used to model the system and calculate these properties\cite{skubicMethodAtomisticSpin2008,nyquistThermalAgitationElectric1928,chandlerIntroductionModernStatistical1987,soderlindPredictionNewEfficient2017,chimataMagnetismUltrafastMagnetization2017,keshavarzExchangeInteractionsCaMnO2017,lochtStandardModelRare2016,yadavHeavymassMagneticModes2018,koumpourasMajorityGateChiral2018}. In the simple case of a bi-linear spin-Hamiltonian, both exchange-coupling parameter $J_{ij}$ and the DMI are needed to correctly describe the studied system, and a common expression is

\begin{equation}
E=-\sum_{i\neq j}J_{ij}\vec{e}_{i}\cdot \vec{e}_{j} -  \sum_{i\neq j}\vec{D}_{ij}\cdot \left( \vec{e}_{i} \times \vec{e}_{j} \right). 
\label{SH}
\end{equation}
In this expression $\vec{e}_{i}$ represents the direction of the magnetic moment at site $i$, and the sum is done over pairs of spins. The interatomic exchange interaction between spins is $J_{ij}$, and theoretical calculations of it and the  DMI, $\vec{D}_{ij}$, offers the possibility to predict important properties, e.g. magnon excitations\cite{skubicMethodAtomisticSpin2008}, which are important for fields such as spintronics and its technological applications. 
However, it was shown that these parameters are strongly affected by non-collinearity in case of non-Heisenberg systems\cite{szilvaTheoryNoncollinearInteractions2017} and, therefore, the properties need to be calculated under a non-collinear formalism. In fact, it has been shown that the calculation of magnon excitations in bcc Fe are closer to the experimental results if one takes into account the calculation of $J_{ij}$ in the non-collinear framework\cite{szilvaInteratomicExchangeInteractions2013}. Furthermore, in more complex systems, DMI can play an important role in the magnon spectra calculation\cite{yadavHeavymassMagneticModes2018,rodriguesFinitetemperatureInteratomicExchange2016}. Also, as the systems get more and more complex, the magnetic structure needs to be taken into consideration for a more precise description of the static and dynamic properties. For that, specifically for the DMI, a non-collinear description is needed. In this work, we show that the DMI is quite different, both regarding the strength and direction, for different magnetic configurations. For instance, for triangular trimers, where the ground state configuration is the N\'eel state (in-plane magnetic moments with an angle of 120$^\circ$ between each other), the DMI is different in direction and strength compared with the DMI calculated considering a collinear configuration. Similar to what the improved $J_{ij}$ could provide in Ref. \citenum{szilvaInteratomicExchangeInteractions2013}, we believe that a general expression for the DMI can lead to new and better understanding of experimental nano-magnetism.

In the absence of SOC interaction, the interatomic exchange coupling can be mapped onto a scalar Heisenberg spin model derived in terms of infinitesimal spin rotations based on Green's function formalism that is known as the Liechtenstein-Katsnelson-Antropov-Gubanov (LKAG) formalism \cite{liechtensteinLocalSpinDensity1987}. The method has been extended for NC atomistic spin configurations in Ref. \citenum{szilvaInteratomicExchangeInteractions2013}. Here, we present a calculation method for the DMI vectors when the SOC is present as a perturbation resulting in (only) anti-symmetric off-diagonal elements of the exchange tensor. This method has two advantages. Firstly, it may give an efficient and expedient estimation for the DMI vectors from a computational point of view. Secondly, this method is able to handle NC atomistic spin configuration by having explicit formulas for the three components for the DMI vectors in the NC framework \cite{szilvaInteratomicExchangeInteractions2013}. We describe here the implementation of this method and illustrate its applicability on selected magnetic nano-sized objects. In order to reach the  established goals, we used the linear muffin-tin orbital - atomic sphere approximation method (RS-LMTO-ASA) to calculate self-consistently the electronic structure for each studied structure.

We have considered Cr triangular trimer supported on the Au(111) surface (Cr$_{3}$/Au(111)) and Mn triangular trimers supported on the Ag(111) (Mn$_{3}$/Ag(111)) and Au(111) (Mn$_{3}$/Au(111)) surfaces. As described in the Methods section, we have employed a linear muffin-tin orbital method in the atomic sphere approximation for the calculation of the electronic structure. Furthermore we considered a real-space version of this method, which enables calculations without periodic boundary conditions, such as clusters of atoms on a substrate. 
The  fcc(111) substrate has been modeled by a slab of 4500 atoms with the experimental lattice parameter of Au (or Au). To simulate the vacuum, outside the surface, we added two layers of empty spheres above the Ag (or Au) surface, in order to provide a basis for the wave function in the vacuum and to treat charge transfers correctly. The calculations of the Mn (or Cr) adsorbed clusters have been   
performed  by embedding the clusters as a perturbation on the previously self-consistently converged
Au(111) or Ag(111) surfaces. The Mn (Cr) sites and the first-nearest-neighbor (NN) atoms (Ag 
or Au, and empty spheres)  around the defect were recalculated self-consistently, while the electronic structure for atoms far from the Mn (Cr) cluster were kept unchanged.
 The collinear and the noncollinear solutions presented here were obtained from fully relativistic calculations, where the spin-orbit interaction was treated at each variational step.\cite{frota-pessoaMagneticBehaviorImpurities2004}
We have performed calculations without  structural relaxation, where the Mn (Cr) atoms occupy the
unrelaxed hollow positions, assuming the experimental lattice parameter of the Ag (or Au) substrate.
The DMI parameters, $D_{ij}$'s, have been computed using an implementation following the theory described in Ref. \citenum{szilvaInteratomicExchangeInteractions2013} and further details are shown in the next section. The obtained values of  $D_{ij}$ are then used to analyse  the effects of non-collinearity onto its strength and direction, aiming to show their differences and consequently their importance for further studies that use them.
%

\par
\section*{Spin interactions}


In the case of pure Heisenberg exchange systems, i.e. when the DMI term in Eqn.1 is neglected, the pairwise energy variation emerges when two spins are simultaneously rotated by infinitesimally small angles, resulting in 
\begin{equation}
\delta E^{H}_{ij} =-2 J_{ij} \delta \vec{e}_{i} \delta \vec{e}_{j},
\label{SH2}
\end{equation}
where the superscript $H$ stands for Heisenberg contribution to the energy of a spin-Hamiltonian, such as Eqn.1. When two spins are rotated at the same time, one has to take into account that the two spins interact both with each other as well as with the rest of the the spin system. The first contribution to the total energy change is denoted by $\delta E^{H}_{ij}$. The same variation term can be calculated for the electronic grand potential, $\delta \Omega^{H}_{ij}$, in terms of multiple scattering electronic structure theory, see Eqn. (\ref{gen22}) in the Methods Section. Comparing $\delta E^{H}_{ij}$ to $\delta \Omega^{H}_{ij}$, the Heisenberg exchange parameter, $J_{ij}$ can be mapped onto $J_{ij}^{*}$, which is defined by Eqn. (\ref{Jstar}) in the Methods section, for a general, NC, case and by Eqn. (\ref{LKAG}) for collinear cases. The later one is commonly referred to as the LKAG formula.   


If the DMI interaction in Eqn.1 is also considered explicitly in the spin-Hamiltonian, the energy of two-site rotations becomes more involved, since one has to add a term to the two-site energy;
\begin{equation}
\delta E^{DM}_{ij}= -2\vec{D}_{ij} \left( \delta \vec{e}_{i} \times \delta \vec{e}_{j} \right). \;
\label{DMH2}
\end{equation}
The superscript $DM$ indicates DMI contributions to the energy of the spin-Hamiltonian. The same mapping procedure can be done as for the Heisenberg term when $\delta E^{DM}_{ij}$ is compared to $\delta \Omega^{DM}_{ij}$ (for details, see the methods section). This leads to the conclusion that $\vec{D}_{ij}$ can be mapped onto $\vec{D}_{ij}^{*}$ which is defined by Eqn. (\ref{DM3}) of the Methods section, for the general, NC, spin-alignment. We show in the Methods section that the presence of SOC interaction mainly contribute to $\delta \Omega^{DM}_{ij}$, see Eqn. (\ref{gen4}).

\begin{figure}[htp]
\begin{center}
\includegraphics[width=0.6\linewidth]{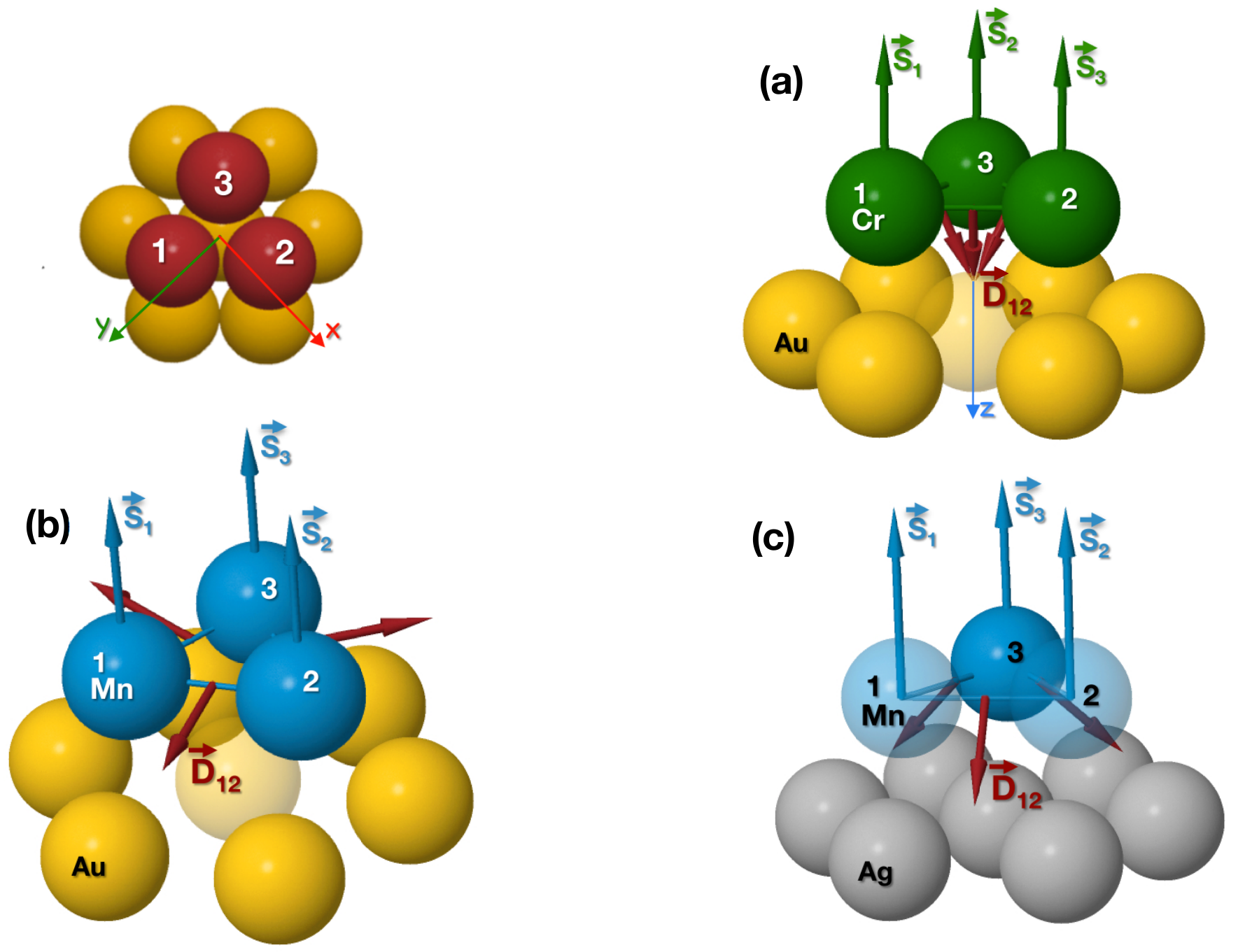}
\caption{\label{figurecol}  Schematic representation of a triangular trimer on an fcc substrate (top left) and the adopted coordinate system. In (a) the moments and DMI vectors of Cr on Au(111) are shown. In b) results for Mn on fcc Au(111) are displayed, and in c) results for Mn on fcc Ag(111) is shown. The dark red arrows denote the DMI direction ($\vec{D}_{ij}$), calculated for a collinear magnetic configuration (ferromagnetic with the magnetic moments  ($\vec{S}_{1}$, $\vec{S}_{2}$, $\vec{S}_{3}$)  perpendicular to the surface along $-\hat{z}$).
}
\end{center}
\end{figure}

\section*{Numerical results}

\subsection*{Cluster geometry and collinear DMI}

In order to test the formalism presented here, we investigate the simple case of a triangular trimer of: (a) Cr on Au(111), Mn on (b) Au(111) and (c) Ag(111) fcc surfaces, considering a collinear magnetic configuration with the atomic moments perpendicular to the surface. For all three cases, the three DMI directions are all related by the $C_{3v}$ symmetry of the triangle trimer, and hence in a accordance to what is expected by Moriya's rules\cite{crepieuxDzyaloshinskyMoriyaInteractions1998} and the surface symmetry. Although, one can see in Fig. \ref{figurecol} that their in-plane and out-of-plane components point to different direction. Concerning their strength, the values are listed in Tab. \ref{table1}. The DMI values are calculated between atoms 1 and 2 (see Fig.~\ref{figurecol}) and, in this particular coordinate system, this DMI vector has two independent components, an in-plane and an out-of-plane, $\vec{D}_{12}=D_\|\, \hat{n} + D_z\, \hat{z}$,  with $\hat{n}=(\hat{x}+\hat{y})/{\sqrt{2}}$.

\begin{table}[h]
\centering
\caption{The strength of the in-plane and out-of-plane components of the DMI vector, in meV. Note that (a), (b) and (c) denotes the systems shown in Fig. \ref{figurecol} of a triangular trimer of Cr on Au(111) surface, Mn on Au(111) and Ag(111) surfaces, respectively. The DMI  calculations were performed for magnetic moments in a ferromagnetic configuration, with the magnetic moments pointing perpendicular to the surface.}
\label{table1}
\begin{tabular}{r|r|r}
  & $D_\|$ & $D_z$ \\
  \hline
(a) Cr on Au(111)   & $-1.51$ & $2.77$ \\ \hline
(b) Mn on Au(111)  & $1.68$ & $-0.04$ \\ \hline
(c) Mn on Ag(111)  & $1.01$ & $0.96$ \\ \hline
\end{tabular}
\end{table}

Note that DMI for the Mn trimer on Ag(111) is comparable with the one of the Mn trimer on Au(111), despite the fact that Au has a larger spin-orbit coupling than Ag. One would therefore expect that an interaction that commonly is ascribed to pure spin-orbit effects, would be larger for the latter substrate. However, the Au 5$d$ bands hybridize only weakly with the 3$d$ states of Mn\cite{belabbesHundRuleDrivenDzyaloshinskiiMoriya2016}. This weakens the influence of the large spin-orbit coupling of the Au atom. In contrast, the hybridization between Mn and Ag is stronger. Although Ag has a weaker spin-orbit strength compared to Au, this increase in hybridization results in a DM interaction that is similar for the two systems. 

If one considers the Heisenberg Hamiltonian for the triangular trimer, for systems where the $J_{ij}$ favours anti-ferromagnetism, which is the case for all of the systems studied here, the magnetic configuration that minimizes the equation is the known N\'eel state, where the magnetic moments are in-plane with an angle of 120$^\circ$ between each other. In that case, the $D_{z}$ component is responsible to lift the degeneracy between different signs of the vector chirality $\vec{\chi}=\vec{S}_1\times\vec{S}_2+\vec{S}_2\times\vec{S}_3+\vec{S}_3\times\vec{S}_1$ in the N\'eel state\cite{antalFirstprinciplesCalculationsSpin2008}. It has been discussed in Ref. \citenum{szilvaInteratomicExchangeInteractions2013} that the Heisenberg picture can be broken and non-Heisenberg quantities can play an important role in case of non-collinearity, therefore changing the values of $J_{ij}$ and $D_{ij}$. In the next section, we show that the calculated DMI's highly depend on their magnetic configuration reference by calculating the DMI's for their respective NC, ground state.

\subsection*{Non-collinear DMI}

\begin{figure}[htp]
\begin{center}
\includegraphics[width=0.6\linewidth]{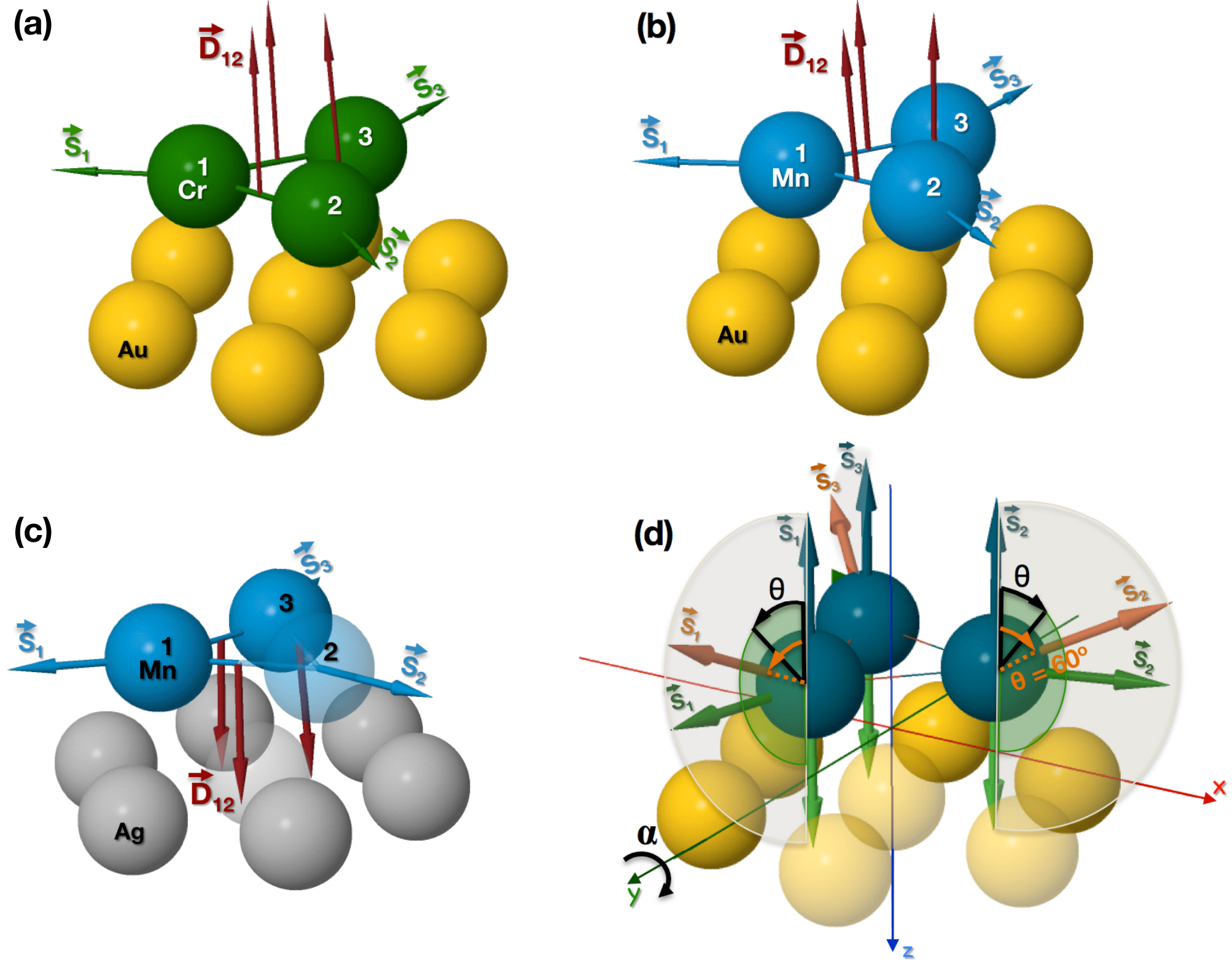}
\caption{\label{figurenoncol2}  Schematic representation of a triangular trimer of: (a) Cr on Au(111), (b) Mn on Au(111) fcc and (c) Mn on Ag(111) fcc surfaces. The dark red arrows denote the DMI directions ($\vec{D}_{ij}$), in case of NC magnetic configuration (N\'eel type state), and they are positioned in-between two atoms which the interaction corresponds to. (d) Diagram with  the different magnetic configurations used in the DMI calculations as a function of the $\theta$ and $\alpha$ angles.}
\end{center}
\end{figure}

At this stage of the calculation, the magnetic moments were allowed to relax in orientation, to find the lowest energy configuration. The co-planar 120$^\circ$ N\'eel 
state was found to be the minimum energy configuration. For this co-planar (CP) configuration, the Heisenberg exchange and DMI interaction was calculated, according to the discussion of the previous section and the Methods section. The results were significantly different, both in direction and strength, compared to the ones found for the collinear case. In the CP case, all three studied systems have the DMI direction almost exclusively perpendicular to the surface, with the in-plane components $D_{x}$ and $D_{y}$ being very weak compared with $D_{z}$. The strengths of these DM interactions are quite intriguing, since they are approximately one order of magnitude larger than the strength of the collinear case.  A similar behaviour was discussed in Ref.\citenum{cardiasDzyaloshinskiiMoriyaInteractionAbsence2020} for the compound Mn$_3$Sn. The comparison between collinear and CP case can be seen in Tab. \ref{table2} and the DMI directions are shown in Fig. \ref{figurenoncol2}. The strength of the DMI is seen to be significantly enhanced for the co-planar orientation. It is unlikely that this enhancement is due to spin-orbit coupling, that is expected to influence the electronic structure in a similar way, for any magnetic configuration. Our explanation of this enhancement, presented below, follows instead an analysis of spin 
currents, that are known to originate from either spin-orbit coupling or 
a co-planarly aligned  magnetism\cite{nordstromNoncollinearIntraatomicMagnetism1996}.

\begin{table}[h]
\centering
\caption{Comparison of the strength of the DMI between the collinear and CP configuration, in meV. Note that (a), (b) and (c) denotes the systems shown in Fig. \ref{figurecol} of a triangular trimer of Cr on Au(111) surface, Mn on Au(111) and Ag(111) surfaces, respectively.}
\label{table2}
\begin{tabular}{r|r|r}
      & $|\vec{D}_{ij}|_\mathrm{collinear}$ & $|\vec{D}_{ij}|_\mathrm{CP}$ \\ \hline
(a) Cr on Au(111)   & $3.16$ & $133.62$ \\ \hline
(b) Mn on Au(111)  & $1.68	$ & $81.74$ \\ \hline
(c) Mn on Ag(111)  & $1.40$ & $59.83$ \\ \hline
\end{tabular}
\end{table}

\subsection*{Influence of spin and charge current on the DMI}

It is known that non-collinearity can lead to spin and charge currents in a system\cite{dossantosdiasChiralitydrivenOrbitalMagnetic2016}. Moreover, the relation between the DMI and spin-current was discussed in Refs. \citenum{kikuchiDzyaloshinskiiMoriyaInteractionConsequence2016} and \citenum{freimuthRelationDzyaloshinskiiMoriyaInteraction2017}. The spin-current that flows between a pair of atoms that the DMI couples, is the main reason for the difference between the collinear and non-collinear (NC) DMI strengths. Also, since both the collinear and the CP N\'eel state have no scalar spin chirality, i.e. $C_{123}=\vec{S}_{1}\cdot(\vec{S}_{2}\times\vec{S}_{3})=0$, the charge-current flowing in the system is weak and is exclusively due to the spin-orbit coupling\cite{dossantosdiasChiralitydrivenOrbitalMagnetic2016}. These mechanisms have been recently discussed in Ref.~\citenum{zhangSpinHallEffect2018}.

In order to analyze in detail the role of the spin and charge currents for the DMI, we first divide the Green's function (GFs) into spin independent $G_{ij}^{0}$ and spin dependent $G_{ij}^{\mu}$ parts. Here $\mu=x,y,z$ and $G_{ij}$ being $9\times 9$ matrices in case of $spd$-orbitals, with orbital indexes written here as $\alpha,\beta$ when needed explicitly as $G_{ij\alpha\beta}$. In this representation, one can describe the behaviour of the GF under a given symmetry operation $T$, that is an operator that transposes simultaneously the orbital and site indices. As discussed previously,\cite{cardiasDzyaloshinskiiMoriyaInteractionAbsence2020} one can further decompose 
the GFs as
\begin{equation}
G_{ij}^{0}=G_{ij}^{00}+G_{ij}^{01}
\label{twoindex}
\end{equation}
and
\begin{equation}
G_{ij}^{\mu}=G_{ij}^{\mu 0}+G_{ij}^{\mu 1},
\label{twoindex2}
\end{equation}
where the second index, introduced in Eqns. \ref{twoindex} and \ref{twoindex2}, denotes whether that part of the GF is is even (0) or odd (1) under $T$. This property is valid for a real basis, in case of spherical harmonics. Further details of this derivation can be seen in Ref.~\citenum{cardiasDzyaloshinskiiMoriyaInteractionAbsence2020}. The definition of the components of the GFs in Eqns.~\ref{twoindex} and \ref{twoindex2} imply that
\begin{equation}
\left(G_{ij \alpha \beta}^{00}\right)^{T}=G_{ji \beta \alpha}^{00}=G_{ij \alpha \beta}^{00},
\end{equation}
and
\begin{equation}
\left(G_{ij \alpha \beta}^{01}\right)^{T}=G_{ji \beta \alpha}^{01}=-G_{ij \alpha \beta}^{01}.
\end{equation}
Furthermore, the two-index formalism can be obtained with the one-index formalism from the relationship

\begin{equation}
G^{0}_{ij}+{G^{0}_{ij}}^{T}=2G^{00}_{ij}
\label{twoidx1}
\end{equation}
\begin{equation}
G^{0}_{ij}-{G^{0}_{ij}}^{T}=2G^{01}_{ij}.
\label{twoidx2}
\end{equation}
The four different parts of the decomposed Green function have direct physical connections,\cite{cardiasDzyaloshinskiiMoriyaInteractionAbsence2020} as tabulated below.

\begin{center}
\begin{tabular}{|l | l|}
\toprule
$G^{00}$ & charge density \\
$G^{01}$ & charge current \\
$G^{\mu 0}$ & spin density \\
$G^{\mu 1}$ & spin current \\
\bottomrule
\end{tabular}
\end{center}
With these relationships, one can rewrite the DMI formula, Eqn.~\ref{DMnew} from the Methods section, as

\begin{equation}
\begin{aligned}
D^{\mu}_{ij}=\frac{1}{2\pi}\Re\int{\mathrm{Tr}\left\{\delta_{i}G_{ij}^{00}\delta_{j}G_{ji}^{\mu 1}+\delta_{i}G_{ij}^{01}\delta_{j}G_{ji}^{\mu 0}\right\}} = \left(D_{ij}^{S}+D_{ij}^{C}\right)^{\mu}.
\label{jij1}
\end{aligned}
\end{equation}
While $G^{00}$ and $G^{\mu 0}$ are the result of the magnetic system, $G^{01}$ and $G^{\mu 1}$ are connected with spontaneous currents that may result from the NC magnetic texture of the system. We will exemplify these mechanism later in the paper. With this formulation, one may divide the DMI in two parts: one related with the spin-current, called $D^{S}\propto G^{\mu 1}$; and other one related with the charge-current $D^{C}\propto G^{01}$. In this interpretation of the DMI, its origin is due to inter-atomic spin and charge currents.

The origin and the mechanism behind the DMI has been intensively discussed in the literature\cite{belabbesHundRuleDrivenDzyaloshinskiiMoriya2016,dossantosdiasChiralitydrivenOrbitalMagnetic2016,kikuchiDzyaloshinskiiMoriyaInteractionConsequence2016,freimuthRelationDzyaloshinskiiMoriyaInteraction2017,karnadModificationDzyaloshinskiiMoriyaInteractionStabilizedDomain2018}. The DMI is known to be a direct effect from the  broken inversion symmetry, caused by the spin-orbit coupling (SOC). In the case analyzed here, a triangle trimer, if the magnetic configuration is collinear, e.g. with the magnetic moments aligned parallel to the surface normal, the SOC induces a current that has direct influence on the DMI. If one turns off the SOC, of this collinear structure, both spin and charge currents vanish, due to the lack of both vector and scalar spin chiralities. However, there are other mechanisms which induce spin and charge currents into a system. The non-collinearity is the relevant mechanism of the presently analyzed systems, from which an additional current flows. Therefore, the currents that may flow in the systems analyzed here, has two different sources which will affect differently the DMI. The electronic structure method used here allows to scale the spin-orbit strength and it is therefore possible to analyse the relative importance of relativistic effects, and the influence of non-collinear magnetic moments. Our results show that in the absence of spin-orbit coupling, the current driven by the non-collinearity is large, with the result that a significant DMI appears, even in the absence of SOC. This has recently also been discussed for Mn$_3$Sn\cite{cardiasDzyaloshinskiiMoriyaInteractionAbsence2020}. A different perspective of these ideas, based on multi-spin and/or multi-scale interaction has recently been discussed by Brinker et al in Ref.~\citenum{brinkerProspectingChiralMultisite2020}. 

In this work, we have performed calculations of the DMI for three different trimers with different magnetic configurations, as represented in Fig.~\ref{figurenoncol2}(d)), namely: (i) a rotation of the magnetic moments from a ferromagnetic configuration to a CP N\'eel magnetic structure and back to a ferromagnetic configuration. This is achieved by rotating the out of plane component of each atomic moment 
from $\theta$ $=$  0$^\circ$ to 180$^\circ$,
while keeping the in-plane component of each magnetic moment at a 120$^\circ$ with respect to the neighbouring Mn (or Cr) atom. We also performed calculations (ii) of the DMI while scaling of the SOC strength, for a configuration when the magnetic moments have an in-plane angle of 120$^\circ$ between each other and an angle of 60$^\circ$ with respect to the $z$-axis ($\theta$ $=$  60$^\circ$). Furthermore, we investigated the effects of a global spin rotation, $R_\alpha$, of the magnetic configuration around the $y$ axis with an angle $\alpha$ (iii). Once the rotation is done, the new DMI was calculated and the vectors were rotated back to the original reference frame.

\begin{figure}[htp]
\begin{center}
\includegraphics[width=0.8\linewidth]{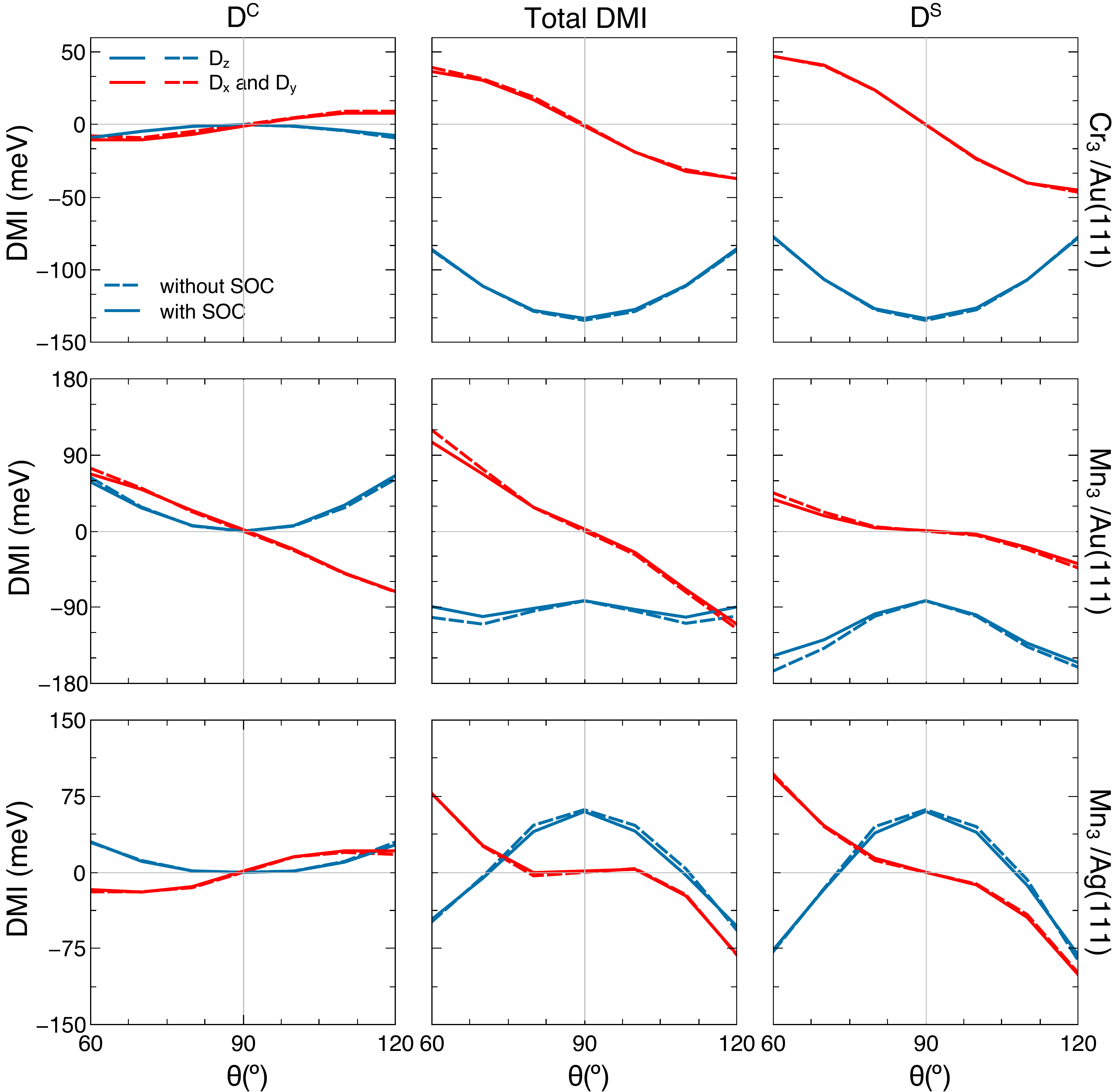}
\caption{\label{figuredm} DMI calculated when varying the vertical angle of the magnetic moments between $\theta=60^\circ$ and $\theta=120^\circ$. On top we have the plot for Cr triangular trimer on Au(111), middle for the triangular trimer of Mn on Au(111) and on bottom the Mn triangular trimer on Ag(111). On the left panels we show results for the DMI part related with the charge-current ($D^{C}$), on the right panels we show data for the spin-current dependent part of the DMI ($D^{S}$), and the middle panels contain the total DMI.  
The blue line denotes the $D_{z}$ component while the red line denotes the in-plane component as $D_{x}=D_{y}=D_\|/\sqrt{2}$. The full line stands for the calculation when the spin-orbit coupling is included, whereas the dashed line denotes the calculation without spin-orbit coupling.}
\end{center}
\end{figure}

In Fig.~\ref{figuredm}, case (i), we show results from a calculation of the DMI for the triangular trimers of Cr on Au(111), Mn on Au(111) and Mn on Ag(111) at every 10$^\circ$, starting from an out-of plane angle of the moment ($\theta$) of 60$^\circ$ and ending at 120$^\circ$. A self consistent calculation for the electronic structure was done for every step, and the DMI was evaluated. The calculations were done including SOC (full line) and without SOC (dashed line), and it was found that in general the difference between the two calculations is minor. Our results also show that if the magnetic configuration is collinear, the DMI only exists in the presence of the SOC (data not shown in Fig.3). However, in the NC 
magnetic configuration case, the DMI is clearly seen from the figure to be non-zero even in the absence of the spin-orbit coupling. This is due to the spin- and charge currents flowing in the system, that are driven by the non-collinearity. 
Note that Fig.~\ref{figuredm} describes the strength of the three components of the Dzyaloshinskii-Moriya interaction. 
It is noteworthy that when $\theta=90^\circ$, i.e. the N\'eel magnetic structure, the scalar spin chirality $C_{123}$ is zero and the charge current flowing in the system is zero. This means that the charge-current dependent part of the DMI, $D^{C}$, vanishes and all the contribution comes from the spin-current dependent $D^{S}$, whose $z$ components are allowed by symmetry. Finally, in the CP N\'eel state and in the lack of SOC, the $x$ and $y$ components of the spin-current dependent part of DMI are also zero.

\begin{figure}[htp]
\begin{center}
\includegraphics[width=0.9\linewidth]{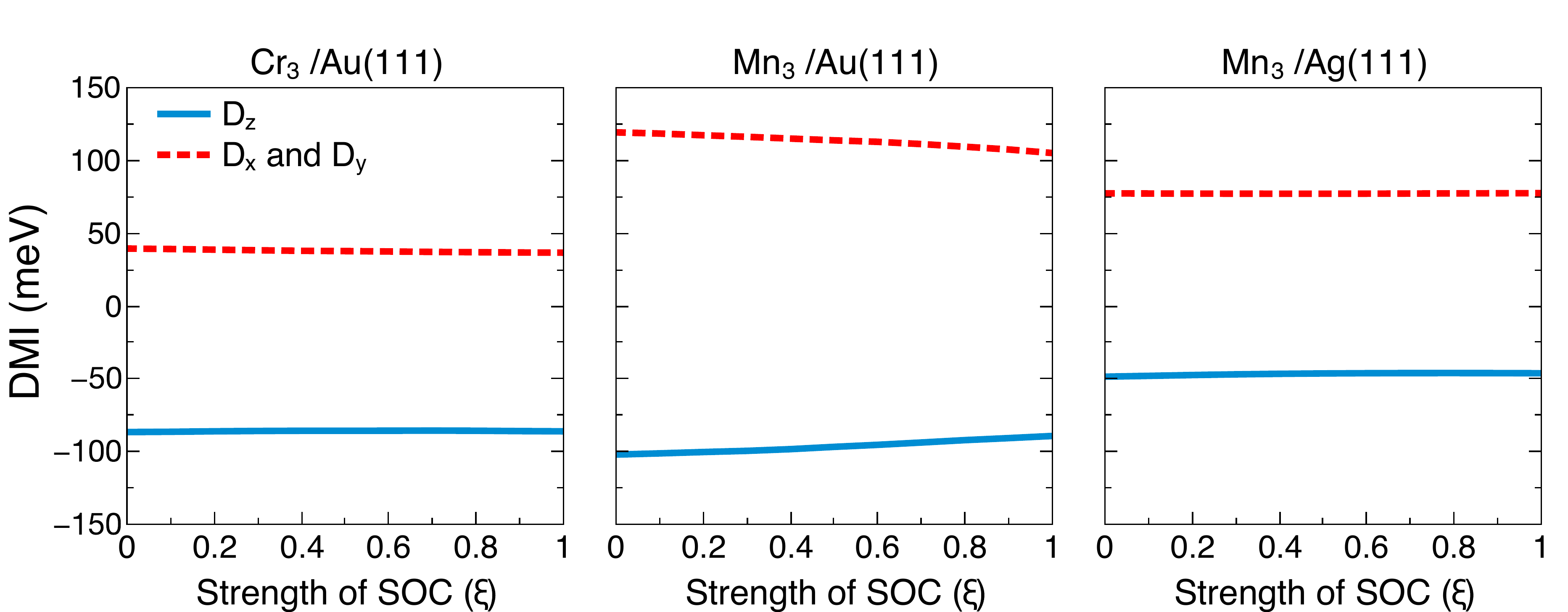}
\caption{\label{socscale} Scaling of the strength of the SOC calculated for a N\'eel like magnetic structure, with a fixed value of angle between atomic moments and the z-axis $\theta=60^\circ$. On the left panel we show results for the Cr triangular trimer on Au(111), the middle panel contains results for the the triangular trimer of Mn on Au(111) and on the right panel we show data for the Mn triangular trimer on Ag(111). The blue  line denotes the $D_{z}$ component while the red line denotes the in-plane components $D_{x}=D_{y}$.}
\end{center}
\end{figure}

In Fig.~\ref{socscale}, case (ii), we show results from a converged calculation, with the magnetic moments having an angle of $\theta=60^\circ$ with respect to the $z$-axis. In this plot we show results of the x-, y- and z-components of the DMI as function of a scaling parameter that was used to tune the strength of the spin-orbit coupling. In this plot, a value of 1 on the abscissa means 100 \% of the true, calculated spin-orbit coupling strength, and e.g. a value 0.5 corresponds to 50 \% of the true, calculated spin-orbit coupling strength. One may note from the figure that for the particular configurations considered, the influence of the SOC is fairly minimal. This demonstrates that the DMI can have a huge, non-relativistic source, that emanates from currents driven by the non-collinearity of the magnetism.

\begin{figure}[htp]
\begin{center}
\includegraphics[width=0.9\linewidth]{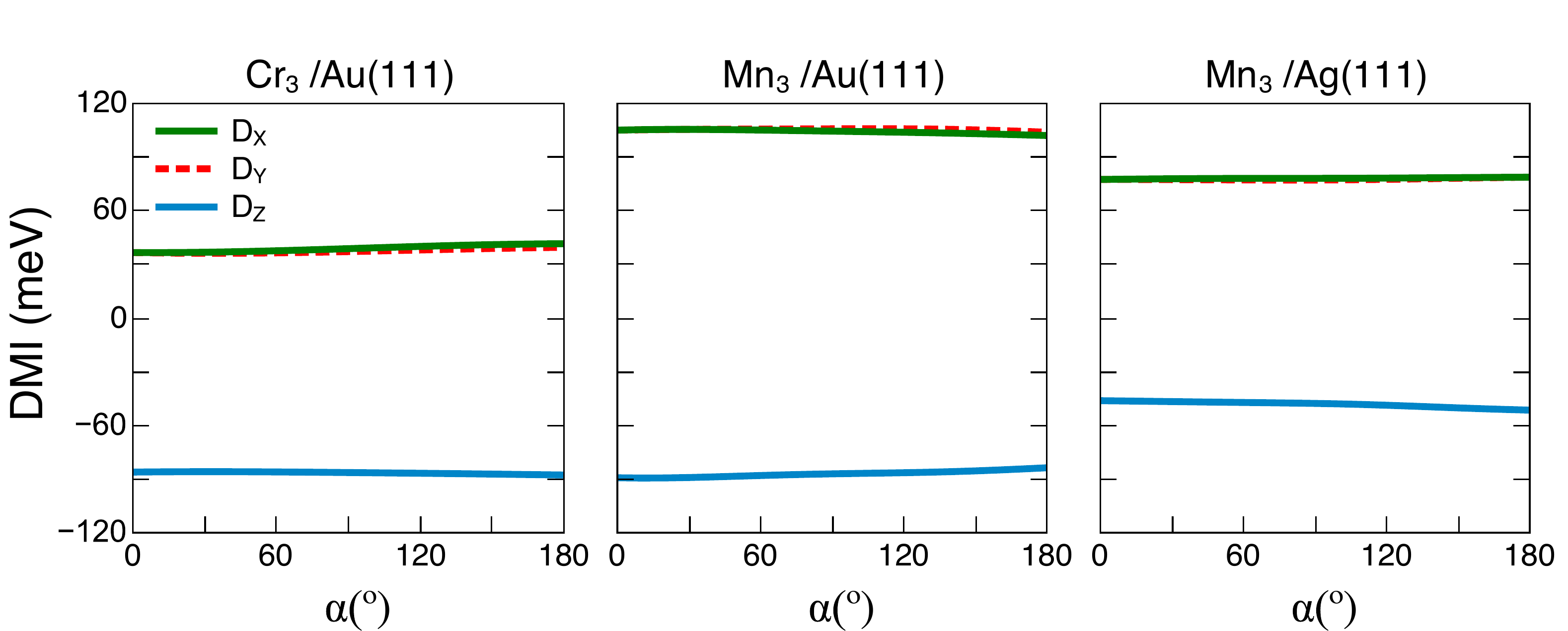}
\caption{\label{rotback}  Calculated DMI when a global spin rotation of the magnetic structure by an angle $\alpha$ around the $y$-axis is performed. On the left panel we show results for the Cr triangular trimer on Au(111), the middle panel contains results for the  triangular trimer of Mn on Au(111) and on the right panel we show data for the Mn triangular trimer on Ag(111).
The green, (dashed) red, and blue lines  denote the $D_{x}$, $D_{y}$, and $D_{z}$ components, respectively.
}
\end{center}
\end{figure}

In Fig.~\ref{rotback}, case (iii), the set-up of Fig.~\ref{socscale} is repeated with $\theta=60^\circ$, but now a global spin rotation, $R_{\alpha}$, of the magnetic moment around the $y$-axis is done. Here, $\alpha$ is the varied angle that is used to rotate moments around the $y$-axis (see Fig.~\ref{figurenoncol2}(d)). The non-relativistic part of the DMI should be a constant and therefore, for zero SOC one expects $\vec{D}_{ij}(\alpha)\cdot R_\alpha \hat{e}_i\times R_\alpha\hat{e}_j=\vec{D}_{ij}(0) \cdot \hat{e}_i\times\hat{e}_j$, for any value of $\alpha$. In the limit of weak SOC, the quantity $R_\alpha^{-1}\vec{D}_{ij}(\alpha)\approx \vec{D}_{ij}(0)$ should be fairly independent of $\alpha$ with any anisotropy directly connected to the SOC. It is possible from Fig.~\ref{rotback} to verify an almost constant value for the DMI with very small dependence on $\alpha$. This corroborates that the DMI is present, in this NC scenario, mostly due to non-relativistic effects, a conclusion also reached in the analysis of Fig.4. 

\section*{Conclusions}

In this work, we have derived the pairwise energy variation between two atoms, in terms of multiple scattering theory, and mapped these results onto a bi-linear Heisenberg Hamiltonian, in order to explicitly evaluate the antisymmetric exchange-coupling parameter, the Dzyaloshinskii-Moriya interaction (DMI), from an implementation in the RS-LMTO-ASA method. The RS-LMTO-ASA method was used to calculate the electronic structure of a Cr triangular trimer on Au(111) surface and Mn triangular trimers on Au(111) and Au(111) surfaces. From these calculations, we have computed the DMI for different magnetic configurations. Firstly, we have found that the DMI evaluated from a ferromagnetic configuration has both strength and direction of the DMI that are in good agreement with previous values found in the literature, when a comparison is possible. Furthermore, we find that, similar to the exchange-coupling parameter, $J_{ij}$, the DMI can strongly depend on the magnetic configuration. To reach this conclusion, the DMI was evaluated from a NC magnetic configuration, for all trimers.
The results show a drastic difference, both for the direction and strength of the DMI, between a collinear and a NC configuration. 


In addition, we argue here that the DMI carries a dependence on the spin and charge currents flowing in the system, that can be induced by either spin-orbit effects or by a NC magnetic configuration. Thus, the fact that the non-collinearity induces a spin and charge current highly contributes to the final value and direction of the DMI. In the particular case of the triangular trimer and the NC magnetic configuration, we have shown that the DMI is mostly influenced by non-relativistic effects. In fact, the results presented here show that the non-relativistic contribution can be orders of magnitude larger than the contribution from spin-orbit coupling. 
Lastly, it is noteworthy that this paper describes a simple way of calculating all the three components of the DMI from a single calculation that works for any considered magnetic configuration.

\section*{Methods}

The fundamental equation of a scalar relativistic multiple scattering theory (MST) formalism is given as
\begin{equation} 
\left( \tau _{ij}^{-1}\right) _{L\sigma ,L^{\prime }\sigma ^{\prime
}}=p_{iL\sigma \sigma ^{\prime }}\delta _{ij}\delta _{LL^{\prime
}}-G^{0}_{ij,LL^{\prime }}\delta _{\sigma \sigma ^{\prime }}\;,
\label{FEMSTT}
\end{equation}%
where $\tau _{ij}$ stands for the scattering path operator (SPO), $\mathbf{p}_{i}$, denotes the inverse of single site scattering operator (ISO). In Eqn. (\ref{FEMSTT}) $L=\left(l,m \right)$ stands for the angular momentum and magnetic quantum numbers, $\sigma$ refers to the spin-index, $\mathbf{G}_{ij}^{0}$ is the free (or bare) electron Green's function and indices $i$ and $j$ refer to the considered lattice sites. $\mathbf{G}_{ij}^{0}$ is calculated from the Hamiltonian of the free particle, hence it is spin-independent. Later on in our presentation, we omit the orbital and spin indices, and the boldface notation stands for quantities in both spin and orbital spaces ($18 \times 18$ matrices in $spd$ basis) while the lack of boldface refers to quantities defined only in the orbital space ($9 \times 9$ matrices in $spd$ basis). We introduce a general notation for the single site scattering operator in a NC framework as
\begin{equation}
\mathbf{t}_{i}(\varepsilon)=t_{i}^{0}(\varepsilon)I_{2}+t_{i}(\varepsilon)\vec{e}_{i} \vec{\sigma} \;,
\label{tdef}
\end{equation}%
where the unit vector $\vec{e}_{i}$ refers to the magnetic spin moment at site $i$ (as it was already defined in the Introduction), $\vec{\sigma}$ stands for the vector formed by Pauli-matrices, $I_{2}$ is the unit matrix in spin space, $t_{i}^{0}$ denotes the non-magnetic (charge) part, and ${t}_{i}$ stands for the magnetic (spin) part of the single site scattering operator, $\varepsilon$ is the energy variable.

For the ISO, one can introduce a similar notation as for the $\mathbf{t}_{i}$ in Eqn. (\ref{tdef}) as follows,
\begin{equation}
\mathbf{p}_{i}=\mathbf{t}_{i}^{-1}=p_{i}^{0}I_{2}+p_{i}\vec{e}_{i} \vec{\sigma} \;.
\label{pdef}
\end{equation}
Later we will need to deal with the variation of the ISO under a small rotation that can be written as
\begin{equation}
\delta \mathbf{p}_{i}=p_{i}\delta \vec{e}_{i}\vec{\sigma}\;, \label{deltapii}
\end{equation}%
where $\delta\vec{e}_{i}$ stands for the deviation of a spin moment after an infinitesimal rotation at site $i$. Finally, the SPO has a structure as
\begin{equation}
\tau _{ij}(\varepsilon )=T_{ij}^{0}I_{2}+\vec{T}_{ij}\vec{\sigma}\;,
\label{deftau3}
\end{equation}%
where $T_{ij}^{0}$ denotes the charge while $\vec{T}_{ij}=\left(T^{x}_{ij}, T^{y}_{ij}, T^{z}_{ij} \right)$ stands for the spin part of the SPO. We have defined the quantities required to calculate the pairwise total energy (grand potential-$\Omega$) variation in a NC framework, i.e., when a spin at site $i$ and an other spin at site $j$ are rotated by an infinitesimal angle simultaneously. As it has been shown in Ref. \citenum{szilvaInteratomicExchangeInteractions2013}, the expression for changes of the grand potential when spins at site $i$ and $j$ are rotated with infinitesimal angles can be written
\begin{equation}
\delta \Omega_{ij} = -\frac{1}{\pi }\int\limits_{-\infty
}^{\varepsilon _{F}}d\varepsilon \operatorname{Im}Tr_{\sigma L}\left( \delta
\mathbf{p}_{i}\tau _{ij}\delta \mathbf{p}_{j}\tau _{ji}\right) \;.  \label{deltae}
\end{equation}%
This expression is commonly evaluated using the Lloyd formula \cite{lloydWavePropagationAssembly1967}. For this purpose, we introduce the matrix 
\begin{equation}
A_{ij}^{\alpha \beta }=\frac{1}{\pi }\int\limits_{-\infty }^{\varepsilon_{F}}d\varepsilon \operatorname{Im} Tr_{L}\left( p_{i}T_{ij}^{\alpha
}p_{j}T_{ji}^{\beta }\right),   \label{AImdeff}
\end{equation}%
and the matrix
\begin{equation}
\hat{A}_{ij}^{\alpha \beta }=\frac{1}{\pi }\int\limits_{-\infty }^{\varepsilon_{F}}d\varepsilon \operatorname{Re} Tr_{L}\left( p_{i}T_{ij}^{\alpha
}p_{j}T_{ji}^{\beta }\right) \;.  \label{ARedeff}
\end{equation}%
Inserting Eqns. (\ref{deltapii}) and (\ref{deftau3}) into Eqn. (\ref{deltae}), and using Eqn. (\ref{AImdeff}), one can formally obtain that
\begin{equation}
\delta \Omega_{ij}=-2 J_{ij}^{*} \delta \vec{e}_{i}\delta \vec{e}
_{j}-4\sum_{\mu ,\nu =x,y,z}\delta e_{i}^{\mu }A_{ij}^{\mu \nu }\delta
e_{j}^{\nu }-2\vec{D}^{*}_{ij}\left( \delta \vec{e}_{i}\times \delta \vec{e}%
_{j}\right) \;, \label{gen22}
\end{equation}%
where expression  
\begin{equation}
J_{ij}^{*}= A_{ij}^{00}-\sum_{\mu =x,y,z}A_{ij}^{\mu \mu}  \;  \label{Jstar}
\end{equation}
and
\begin{equation}
D_{ij}^{*\mu}= \hat{A}_{ij}^{0 \mu}- \hat{A}_{ij}^{\mu 0} \;  \label{DM3}
\end{equation}
have been introduced. The first term in Eqn. (\ref{gen22}) is denoted $\delta \Omega_{ij}^{H}$ (for Heisenberg like contributions) while the third term is denoted $\delta \Omega_{ij}^{DM}$ (for DM like controbutions). Here, the new superscript $*$ was introduced in order to differentiate the terms derived from the MST from the ones presented in Eqn.~\ref{SH}. However, a practical calculation of exchange parameters used in an effective spin-Hamiltonian, amount to identifying the connection between the parameters in Eqn.1 and those in Eqns.\ref{Jstar} and \ref{DM3}.

We analyse in this paragraph the case when the atomic moments are in a collinear arrangement (and the spin orbit-interaction is absent). One can then write that $\vec{T}_{ij}=\left((0, 0, T^{z}_{ij} \right)$, and the component of the SPO for the up and down spin channels can be defined as $T_{ij}^{\uparrow}=T_{ij}^{0}+T_{ij}^{z}$ while $T_{ij}^{\downarrow}=T_{ij}^{0}-T_{ij}^{z}$. It can be shown that $D_{ij}^{*\mu}=0$ in this case and the second term in Eqn. (\ref{gen22}) gives only higher (fourth) order contribution to $\Omega$ in terms of the angle variation. Only the first term is left in which $J_{ij}^{*}$ is simplified to 
\begin{equation}
J_{ij}^{*}= A_{ij}^{00}-A_{ij}^{zz}= A_{ij}^{\uparrow \downarrow} \;.  \label{LKAG}
\end{equation}
The expression defined in Eqn. (\ref{LKAG}) is widely used for ab-initio calculations of Heisenberg exchange, and is commonly referred to as the LKAG formula. 

In the rest of the Section, we analyse for a general non-collinear magnetic configuration
what contributions are added to the variation of $\Omega$, when the SOC is considered as a perturbation, i.e. only the terms that being of first order in the spin-orbit coupling parameter, $\xi$, are kept. One can then write that
\begin{equation}
\delta \Omega_{ij}^ {\prime} = -\frac{1}{\pi }\int\limits_{-\infty
}^{\varepsilon _{F}}d\varepsilon \operatorname{Im}Tr_{\sigma L}\left( \delta
\mathbf{P}^{\prime}_{i}\tau^{\prime}_{ij}\delta \mathbf{P}^{\prime}_{j}\tau^{\prime}_{ji}\right) \;,  \label{deltae2}
\end{equation}%
where the perturbed quantities are denoted by the symbol $\prime$. Our task now is to determine the $\delta \mathbf{P}^{\prime} $-s and $\tau^{\prime}$-s.
First, we determined the perturbed single scattering matrix with the help of perturbed Green function. Next, we introduce the perturbed $\mathbf{G}^{\prime}_{ij}$ where $\mathbf{G}^{\prime}_{ij}=\mathbf{G}^{0}_{ij}+\Delta \mathbf{G}_{ij}$. The matrix $\mathbf{G}^{0}_{ij}$ has the structure of $G_{ij}^{0}I_{2}$, while $\Delta \mathbf{G}_{ij}$ can be written as $\xi \vec{\Gamma}_{ij} \vec{\sigma}$, where its vector component $\Gamma_{ij}^{\mu}$ ($\mu$= $x$, $y$ and $z$) is obtained as
\begin{equation}
\Gamma_{ij}^{\mu}= \sum_{k} \mathbf{G}^{0}_{ik} L^{\mu}  \mathbf{G}^{0}_{kj},  \;
\label{Gammaprop}
\end{equation}%
and where $L^{\mu}$ is a component of the angular momentum operator. This implies that $\vec{\Gamma}_{ij}$ transforms under T as
\begin{equation}
\left(\Gamma_{ij}^{\mu}\right)^{T}=-\Gamma_{ji}^{\mu}\;.
\label{Gammaprop2}
\end{equation}%
Keeping the leading terms, we get that $\mathbf{t}^{\prime}_{i}=\mathbf{t}_{i}+\xi \mathbf{t}_{i} \left(\vec{\Gamma}_{ii} \vec{\sigma}\right)  \mathbf{t}_{i}$ which can alternatively be written as $\mathbf{t}^{\prime}_{i}=\mathbf{t}_{i} \left(I+\xi  \left(\vec{\Gamma}_{ii} \vec{\sigma}\right)  \mathbf{t}_{i} \right)$ where $I$ is the unit matrix. Then $\left(\mathbf{t}^{\prime}_{i} \right)^{-1}= \mathbf{t}_{i}^{-1}- \xi \left(\vec{\Gamma}_{ii} \vec{\sigma}\right)$ and 
\begin{equation}
\mathbf{P}_{i}^{\prime} \simeq \mathbf{P}_{i} - \xi \left(\vec{\Gamma}_{ii} \vec{\sigma}\right)=p_{i}^{0}+ \left(p_{i} \vec{e}_{i} - \xi \vec{\Gamma}_{ii} \right) \vec{\sigma} \;,
\end{equation}%
see Eqns. (\ref{tdef}-\ref{pdef}). This implies that
\begin{equation}
\delta \mathbf{P}_{i}^{\prime} \simeq p_{i} \delta \vec{e}_{i} \vec{\sigma}= \delta \mathbf{P}_{i} \;,
\end{equation}%
i.e., we have to calculate the simpler expression
\begin{equation}
\delta \Omega_{ij}^ {\prime} \simeq -\frac{1}{\pi }\int\limits_{-\infty
}^{\varepsilon _{F}}d\varepsilon \operatorname{Im}Tr_{\sigma L}\left( \delta
\mathbf{P}_{i}\tau^{\prime}_{ij}\delta \mathbf{P}_{j}\tau^{\prime}_{ji}\right) \;  \label{deltae3}
\end{equation}%
that contains the perturbed SPO, however, the ISO is present in a non-perturbed form.

Next we need to determine the perturbed SPO. It can be shown that
\begin{equation}
\tau^{\prime}_{ij}=\tau_{ij}+ \Delta \tau_{ij} \;,
\end{equation}%
where
\begin{equation}
\begin{aligned}
\Delta \tau_{ij} = \xi \mathbf{t}_{i}  \left(\vec{\Gamma}_{ii} \vec{\sigma}\right) \mathbf{t}_{i} \delta_{ij}+\xi \mathbf{t}_{i}  \left(\vec{\Gamma}_{ii} \vec{\sigma}\right) \mathbf{t}_{i} \mathbf{G}_{ij}^{0} \mathbf{t}_{j}+\xi \mathbf{t}_{i} \mathbf{G}_{ij}^{0} \mathbf{t}_{j} \left(\vec{\Gamma}_{jj} \vec{\sigma}\right) \mathbf{t}_{j} + \xi \mathbf{t}_{i}  \left(\vec{\Gamma}_{ij} \vec{\sigma}\right) \mathbf{t}_{j} \;.
\label{deltalong}
\end{aligned}
\end{equation}%
It can furthermore be demonstrated that $\Delta \tau_{ij}$ has the same structure as $\tau_{ij}$, i.e.,
\begin{equation}
\Delta \tau_{ij} = \Delta T^{0}_{ij} I_{2} + \Delta \vec{T}_{ij} \vec{\sigma} \;.
\end{equation}%
Using Eqns. (\ref{Gammaprop}) and (\ref{deltalong}), it can be shown by using Eqn. (\ref{Gammaprop2}) that
\begin{equation}
\left(\Delta T_{ij}^{\alpha}\right)^{T}=-\Delta T_{ji}^{\alpha}\;,
\label{Dtauprop}
\end{equation}
where $\alpha$ runs over the $0$, $x$, $y$ and $z$. This implies that
\begin{equation}
Tr_{L}\left( p_{i} \Delta T^{\alpha}_{ij} p_{j} T_{ji}^{\beta}\right)= - Tr_{L}\left( p_{i}  T^{\beta}_{ij} p_{j} \Delta T_{ji}^{\alpha}\right) \;
\label{first}
\end{equation}%
and
\begin{equation}
Tr_{L}\left( p_{i}  T^{\alpha}_{ij}  p_{j} \Delta T_{ji}^{\beta}\right)= - Tr_{L}\left( p_{i}  \Delta T^{\beta}_{ij} p_{j} T_{ji}^{\alpha}\right) \;.
\label{second}
\end{equation}
Next, we define the matrices $A^{\prime \alpha \beta }_{ij}$ and $\hat{A}_{ij}^{\prime \alpha \beta }$ as follows: we replace $T^{\alpha}_{ij}$ by $T^{\prime \alpha}_{ij}$ and $T^{\beta}_{ji}$ by $T^{\prime \beta}_{ji}$ in Eqns. (\ref{AImdeff}) and (\ref{ARedeff}), respectively,  where $T^{\prime \alpha}_{ij}=T^{\alpha}_{ij}+ \Delta T^{\alpha}_{ij} $ and $T^{\prime \beta}_{ji}=T^{\beta}_{ji}+ \Delta T^{\beta}_{ji} $. Finally, we define $\vec{D'}^{*}_{ij}$ as follows,
\begin{equation}
D_{ij}^{'*\mu}= \hat{A}_{ij}^{\prime 0 \mu}- \hat{A}_{ij}^{\prime \mu 0} \;.  \label{DMnew}
\end{equation}
By using these definitions, we get from Eqn. (\ref{deltae3}) that 
\begin{equation}
\begin{aligned}
\delta \Omega_{ij}^{\prime}=-2\left( A_{ij}^{\prime 00}-\sum_{\mu =x,y,z}A_{ij}^{ \prime \mu \mu
}\right) \delta \vec{e}_{i}\delta \vec{e}
_{j}-2\sum_{\mu ,\nu =x,y,z}\delta e_{i}^{\mu } \left( A_{ij}^{\prime \mu \nu} + A_{ij}^{\prime \nu \mu} \right) \delta
e_{j}^{\nu }-2\vec{D'}^{*}_{ij}\left( \delta \vec{e}_{i}\times \delta \vec{e}%
_{j}\right) \;,  \label{gen3}
\end{aligned}
\end{equation}%
where $A^{\prime \alpha\beta}_{ij} \neq A^{\prime \beta \alpha}_{ij} $. However, by using Eqns. (\ref{first}) and (\ref{second}), it can be shown that $A^{\prime \alpha \alpha}_{ij} = A^{\alpha \alpha}_{ij} $. In addition, by using the same equations, it can be also written that $A_{ij}^{\prime \mu \nu} + A_{ij}^{\prime \nu \mu} = 2 A_{ij}^{ \mu \nu}$. This implies that Eqn. (\ref{gen3}) will be reduced to the expression
\begin{equation}
\begin{aligned}
\delta \Omega_{ij}^{\prime}=-2 J_{ij}^{*} \delta \vec{e}_{i}\delta \vec{e}
_{j}-4\sum_{\mu ,\nu =x,y,z}\delta e_{i}^{\mu }A_{ij}^{\mu \nu }\delta
e_{j}^{\nu }-2\vec{D'}^{*}_{ij}\left( \delta \vec{e}_{i}\times \delta \vec{e}%
_{j}\right) \;. \label{gen4}
\end{aligned}
\end{equation}%
From this expression we note that primed quantities, that signify that they contain linear contributions of the spin-orbit coupling, are only found for the DM interaction. Hence, we conclude from perturbation theory that spin-orbit interaction influences only the DM interaction.

Summarizing the methods section, we have shown that the exchange parameter $J_{ij}$ defined in the spin model, Eqn. (\ref{SH}), can mapped onto $J_{ij}^{*}$ given by Eq. (\ref{Jstar}) for a general, NC, spin arrangement. Note that the mapping is correct if the second term in Egn. (\ref{gen22}) is not relevant. This is the case for instance with symmetry resolved exchange parameter in the $T_{2g}$ channel in bcc Fe bulk\cite{szilvaTheoryNoncollinearInteractions2017}. We can also note that the $J_{ij}$ parameter in Eqn. (\ref{SH}) for collinear magnets is the LKAG parameter given by Eqn. (\ref{LKAG}). In addition, we have shown that the DMI vector $\vec{D}_{ij}$ can be calculated from the electronic structure by the $\vec{D}^{*}_{ij}$ formula given by Eqn. (\ref{DM3}). This formula is general, and holds for any kind of NC spin configuration. We have also shown, see Eqn. (\ref{gen4}) that the SOC interaction contributes only to the DM interaction in leading order.



\subsection*{Computational details}
The calculations were performed using the {\it ab initio}  RS-LMTO-ASA method\cite{frota-pessoaFirstprinciplesRealspaceLinearmuffintinorbital1992,klautauOrbitalMoments3d2005,bergmanMagneticInteractionsMn2006,bergmanMagneticStructuresSmall2007,bezerra-netoComplexMagneticStructure2013,klautauMagneticPropertiesCo2004,frota-pessoaInfluenceInterfaceMixing2002}, which is suitable to describe the physics of isolated clusters supported on surfaces in an efficient way, since it is real-space based and does not depend on translational symmetry. Moreover, the RS-LMTO-ASA method has been generalized  to describe non-collinear (NC) magnetism\cite{frota-pessoaFirstprinciplesRealspaceLinearmuffintinorbital1992,bergmanMagneticStructuresSmall2007,bergmanNoncollinearMagnetisationClusters2006},  and is based on the Haydock recursion method\cite{haydockRecursiveSolutionSchrodinger1980}. Our Hamiltonians are constructed within an RS-LMTO-ASA formalism\cite{andersenLinearMethodsBand1975}, and therefore, all calculations presented here are fully self-consistent, and the spin densities were treated within the local spin-density approximation (LSDA)\cite{barthLocalExchangecorrelationPotential1972}. The continued fraction, that occurs in recursion method,  have been terminated  with the Beer-Pettifor terminator\cite{beerRecursionMethodEstimation1984} after 21 recursion levels.



\bibliography{dmipaper}

\section*{Acknowledgements}
R. C. and A. B. K. acknowledge financial support from CAPES and CNPq, Brazil. 
The calculations were performed at the computational facilities of the CENAPAD at University of Campinas, SP, Brazil and the Swedish National Infrastructure for Computing (SNIC). O.E. acknowledges support from the Knut and Alice Wallenberg foundation, the foundation for strategic research (SSF), the Swedish Energy Agency (Energimyndigheten), STandUPP and the Swedish Research Council (VR). L.N. and Y.K. also acknowledge the support from VR. Finally, we would like to acknowledge the support from eSSENCE.

\section*{Author Contributions}
R.C., M.M.B-N., M.S.R., A.B., A.B.K., and L.N. conceived and designed these studies.  L.N., A.Sz., J.F., and  Y.K. contributed  to  the  model  development. R.C. and A. B. performed the \textit{ab initio} calculations. The first version of the manuscript was prepared by R.C., A.B.K. and O.E. All authors discussed the results and contributed to preparing the paper. 

\section*{Additional information}
\textbf{Competing Interests}: The authors declare that they have no competing interests.

\end{document}